\documentclass[showpacs,aps,superscriptaddress,prb,twocolumn,amsmath]
{revtex4}
\usepackage{txfonts}
\usepackage{color}
\usepackage{amssymb}
\usepackage{graphicx}
\usepackage{dcolumn}
\usepackage{bm}
\usepackage{hyperref}
\usepackage{color}

\begin{document}
\preprint{APS/123-QED}
\title{New allotropes of phosphorene with remarkable stability and intrinsic piezoelectricity}
\author{Zhenqing Li}
\affiliation{Hunan Key Laboratory for Micro-Nano Energy Materials
and Devices, Xiangtan University, Hunan 411105, P. R. China;}
\affiliation{School of Physics and Optoelectronics, Xiangtan
University, Xiangtan 411105, China.}
\author{Chaoyu He}
\email{hechaoyu@xtu.edu.cn}\affiliation{Hunan Key Laboratory for Micro-Nano Energy Materials
and Devices, Xiangtan University, Hunan 411105, P. R. China;}
\affiliation{School of Physics and Optoelectronics, Xiangtan
University, Xiangtan 411105, China.}
\author{Tao Ouyang}
\affiliation{Hunan Key Laboratory for Micro-Nano Energy Materials
and Devices, Xiangtan University, Hunan 411105, P. R. China;}
\affiliation{School of Physics and Optoelectronics, Xiangtan
University, Xiangtan 411105, China.}
\author{Chunxiao Zhang}
\affiliation{Hunan Key Laboratory for Micro-Nano Energy Materials
and Devices, Xiangtan University, Hunan 411105, P. R. China;}
\affiliation{School of Physics and Optoelectronics, Xiangtan
University, Xiangtan 411105, China.}
\author{Cao Tang}
\email{C$_$Tang@xtu.edu.cn}\affiliation{Hunan Key Laboratory for Micro-Nano Energy Materials
and Devices, Xiangtan University, Hunan 411105, P. R. China;}
\affiliation{School of Physics and Optoelectronics, Xiangtan
University, Xiangtan 411105, China.}
\author{Rudolf A. R\"{o}mer}
\affiliation{Department of physics, the university of warwick coventry, CV4 8UW, UK.}
\author{Jianxin Zhong}
\email{jxzhong@xtu.edu.cn}\affiliation{Hunan Key Laboratory for
Micro-Nano Energy Materials and Devices, Xiangtan University, Hunan
411105, P. R. China;} \affiliation{School of Physics and
Optoelectronics, Xiangtan University, Xiangtan 411105, China.}
\date{\today}
\pacs{73.20.At, 61.46.-w, 73.22.-f, 73.61.Cw}
\begin{abstract}
In this letter, we show that a new class of two-dimensional phosphorus allotropes can be constructed via assembling the previously proposed ultrathin metastable phosphorus nanotube into planar structures in different stacking orientations. Based on first-principles method, the structures, stabilities and fundamental electronic properties of these new two-dimensional phosphorus allotropes are systematically investigated. These two-dimensional phosphorus allotropes possess remarkable stabilities due to the strong inter-tube van der Waals interactions, which cause an energy release of about 30-70 meV/atom depending on their stacking manners. Our results show that most of these two-dimensional van der Waals phosphorene allotropes are energetically more favorable than the experimentally viable black $\alpha$-P and blue $\beta$-P. Three of them showing relatively higher probability to be synthesized in future are further confirmed to be dynamically stable semiconductors with strain-tunable band gaps and remarkable piezoelectricity, which may have potential applications in nano-sized sensors, piezotronics, and energy harvesting in portable electronic nano-devices.\\
\end{abstract}
\maketitle
\indent Black phosphorene ($\alpha$-P) is considered as a formidable competitor to graphene and other two-dimensional materials for applications in future nano-electronics due to its significant band gap \cite{3, 4} and high carrier mobility\cite{2}. It was successfully exfoliated from its three-dimensional counterpart in 2014 \cite{1, 2} for the first time and now can be synthesized through low-cost and high-yield methods, such as liquid exfoliation \cite{6x} and chemical deposition methods \cite{7x, 8x}. Blue phosphorene ($\beta$-P), another two-dimensional phosphorus allotrope, was firstly proposed by Zhu et.al based on first-principles calculations \cite{5} in 2014 and recently synthesized by Zhang et.al through epitaxial growth method \cite{blueE}. Progress in synthesis of two-dimensional materials has stimulated significant interests in exploring new two-dimensional phosphorus materials. \\
\indent Considered that the possible crystal structures for two-dimensional phosphorus synthesized from deposition methods or epitaxial growth methods always depend on the substrate, pressure, temperature and other experimental conditions, many other potential \cite{7, 8, 9, 10, 101, 1x, 121, 122} candidates beyond black $\alpha$-P and blue $\beta$-P have been theoretically proposed and are expected to be synthesized in future. For example, the atomic thing layers of $\gamma$-P \cite{7}, $\delta$-P \cite{7}, $\theta$$_0$-P \cite{8} and red phosphorene \cite{9}, the diatomic thin layers $\eta$-P \cite{10}, $\theta$-P \cite{10} and their transformations \cite{101} (G1, G2, B1, B2 and B3) with pentagons, the multi-atomic thin layer of Hittorfene \cite{1x}, as well as the porous phosphorene allotropes were previously proposed \cite{121,122} through the topological modeling method. Recently, a new Helical Coil phosphorus was theoretically predicted \cite{ptt} and its corresponding allotropes was experimentally synthesized \cite{pte} in carbon nanotube reactor.\\
\begin{figure}
\center
\includegraphics[width=3.5in]{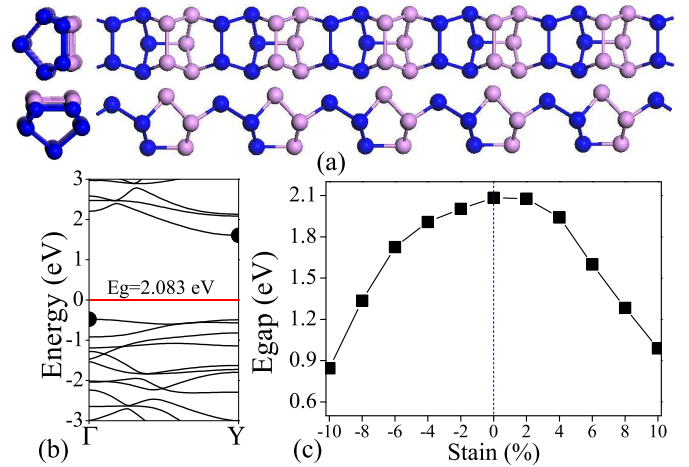}\\
\caption{ Crystalline views of the um-PNT from different directions (a). The calculated electron band structures (b) and strain-dependent band gaps (c) of the um-PNT.}\label{fig1}
\end{figure}
\begin{figure*}
\center
\includegraphics[width=6.0in]{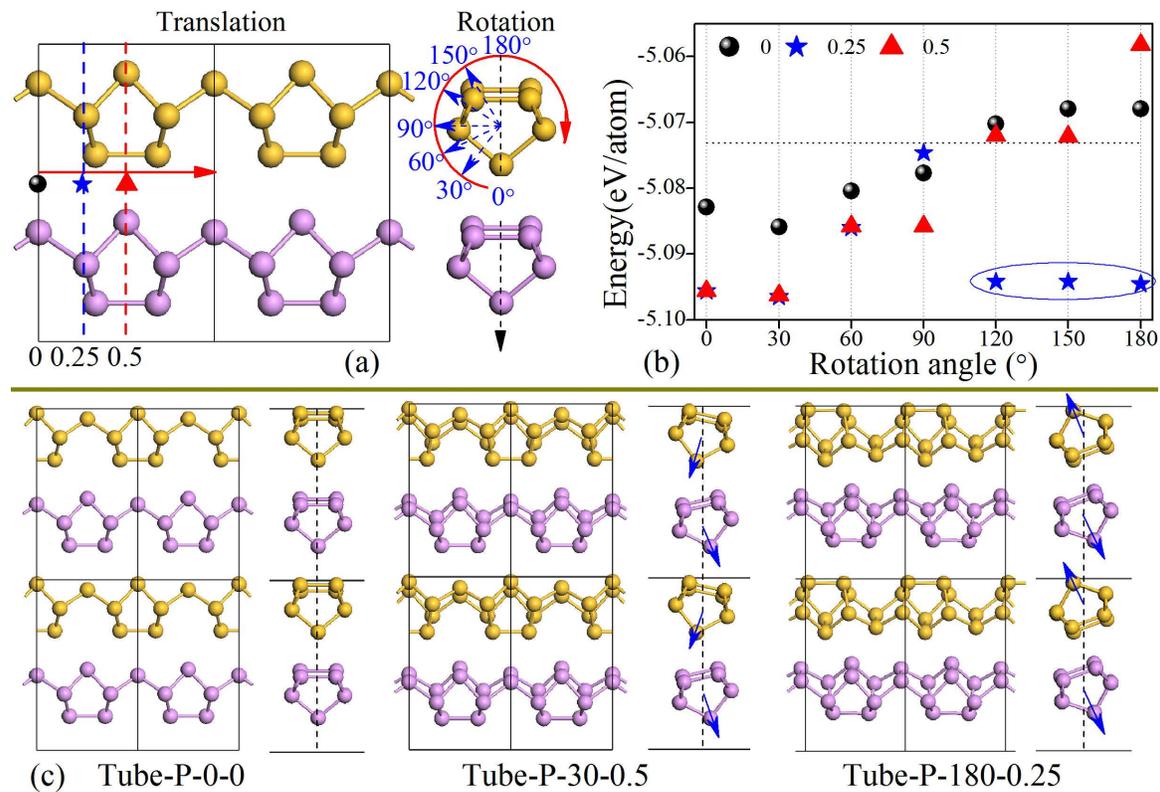}\\
\caption{Sketch map of the translation and self-rotation for assembling one-dimensional um-PNT into two-dimensional phosphorus allotropes (a). The calculated average energy for the 21 new two-dimensional van der Waals phosphorene allotropes (b). The crystalline views of the most stable three ones, Tube-P-0-0.5, Tube-P-30-0.5 and Tube-P-180-0.25 (c).}\label{fig2}
\end{figure*}
\indent Phosphorene allotropes with lone pairs on the surface like SnS and SnSe \cite{Pi1, Pi2} are expected to be excellent two-dimensional piezoelectric materials\cite{Pi3, Pi4, Pi5, Pi6}. However, the experimental viable black $\beta$-P and blue $\alpha$-P, as well as most of all the previously proposed two-dimensional allotropes of phosphorus are centrosymmetric and consequently non-polar, which indicate that they are non-piezoelectric. In this letter three dynamically stable two-dimensional van der Waals phosphorus crystals are obtained via planer stacking of the previously proposed ultrathin metastable phosphorus nanotube (um-PNT, P$_\infty$\cite{ptt}) and confirmed energetically more favorable than black $\alpha$-P and blue $\beta$-P. Our results show that these three new phosphorene allotropes are excellent candidates for potential application in nano-electronics according to their strain-turnable band gaps and remarkable piezoelectricity. We expect future efforts will be paid to synthesize them due to their remarkable energetic stabilities. \\
\indent Our calculations were carried out by using the density functional theory (DFT) within generalized gradient approximations (GGA)\cite{12} as implemented in Vienna ab initio simulation package (VASP)\cite{13, 14}. The interactions between nucleus and the 3s$^2$3p$^3$ valence electrons of phosphorus atoms were described by the projector augmented wave (PAW) method\cite{15, 16}. To ensure the accuracy of our calculations, a plane-wave basis with a cutoff energy of 500 eV was used to expand the wave functions and the Brillouin Zone (BZ) sample meshes were set to be dense enough (less than 0.21 $\AA$$^{-1}$) for each system. The structures of all phosphorene allotropes were fully optimized up to the residual force on every atom less than 0.001 eV/$\AA$ before property investigations. The optimized exchange van der Waals functional (optB86-vdW)\cite{17, 18} was also applied to take into account van der Waals interactions in the systems. The vibrational properties of the three new phosphorene allotropes more favorable than black $\alpha$-P and blue $\beta$-P were investigated through the PHONON package\cite{20} with the forces calculated from VASP to confirm their dynamical stabilities. \\
\indent The crystal structure of the um-PNT consists of phosphorus pentagons is shown in Fig.1 (a). It is structurally very similar to the fundamental structure-units in violet phosphorus and red phosphorus \cite{v1,1x,r1}. Our calculated results show that it is of about 44 meV/atom higher than the black $\alpha$-P in energy. And it is more favorable than most of the previously proposed two-dimensional phosphorene allotropes. Especially, um-PNT is also more stable than most of the previously proposed a-PNTs and z-PNTs \cite{PNT}. The smallest phosphorus tube proposed before is the a-PNT with tube radius of about 2.3 $\AA$, which is of about 100 meV/atom higher than black phosphorene and less stable than our um-PNT. The radius of um-PNT is only 1.6 $\AA$. Thus, um-PNT can be considered as the smallest phosphorus tube so far. We then investigated the fundamental electronic property of such an um-PNT. The calculated band structure shown in Fig.1 (b) indicates that um-PNT is a semiconductor with indirect band gap of about 2.02 eV. Further investigation shows that um-PNT is a good candidate for one-dimensional semiconductor applications in view of that its band gap can be effectively modulated by strains.\\
\indent As shown in Fig.1 (a), two um-PNTs are arranged in an orthorhombic cell as a starting structure, in which the violet tube is fixed and the yellow one is free for operation. Based on the well defined operators T$_i$ (Translation, i=0, 0.25 and 0.5 \textbf{b}) and R$_j$ (Self-rotation, j=0$^\circ$, 30$^\circ$, 60$^\circ$, 90$^\circ$, 120$^\circ$, 150$^\circ$ and 180$^\circ$) as indicated in Fig.1 (a), 21 initial structures can be constructed. We name them as Tube-P-j-i according to the operation $\{$R$_j$,T$_i$$\}$ applied on the free yellow tube, where j means the rotation angle and i means the translation. For example, Tube-P-180-0.25 means the mutation of the initial structure Tube-P-0-0 through a self-rotation of 180$^\circ$ and a translation of 0.25 \textbf{b} along the tube (Y direction) on the yellow tube. \\
\indent We then optimized the crystalline structures of these 21 new phosphorus allotropes through first-principles calculations. The calculated total energies of black $\alpha$-P and the 21 new two-dimensional van der Waals phosphorus allotropes are shown in Fig.2 (b). According to our results, all these new two-dimensional phosphorus allotropes are more stable than the one-dimensional um-PNT in energy of about 30-70 meV/atom dependent on the assembling manner. With the release of remarkable energy after assembling into planar array, most of these new two-dimensional phosphorus allotropes are energetically more stable than the viable black $\alpha$-P and blue $\beta$-P. From the results in Fig.2 (b), we can see that some initial structures are degenerate in energy after optimization. We checked the optimized structures and found that Tube-P-0-0.25, Tube-P-30-0.25 and Tube-P-60-0.25 are degenerate to Tube-P-0-0.5, Tube-P-30-0.5 and Tube-P-60-0.5, respectively. Both Tube-P-120-0.25 and Tube-P-150-0.25 are degenerate to Tube-P-180-0.25. We further confirmed that the most stable three ones are Tube-P-0-0.5, Tube-P-30-0.5 and Tube-P-180-0.25, whose energies are of about 22 meV/atom, 23 meV/atom and 21 meV/atom lower than that of the black phosphorene. The crystal structures of the most stable three ones are shown in Fig.2 (c), where we can see that they are obviously different to each other. We also found that Tube-P-0-0.5 keeps its initial structure well in the optimization process, but Tube-P-30-0.5 and Tube-P-180-0.25 have obvious changes in orientation angle as indicated in Fig.2 (c).\\
\indent From the view of thermodynamics, a structure with lower energy generally means higher probability to be synthesized in experiments if it is dynamically possible. Tube-P-0-0.5, Tube-P-30-0.5 and Tube-P-180-0.25 with remarkable stability exceeding black $\alpha$-P are expected to be synthesized in future using vapor deposition methods. The dynamical stabilities of these three new phosphorene allotropes are thus needed to be confirmed the possibility of synthesis. We studied their vibrational properties through PHONON package combined with VASP. The results are shown in Fig.3. We can see that the phonon band structures of Tube-P-0-0.5, Tube-P-30-0.5 and Tube-P-180-0.25 are free of soft modes associated with structural instability. We also examined the whole Brillouin Zone and found no imaginary states in their phonon density of states. The results show that allotrope Tube-P-0-0.5, Tube-P-30-0.5 and Tube-P-180-0.25 are dynamically viable too. Based on these vibrational spectra, we calculated the Helmholtz free energies and compared them with that of the black $\alpha$-P. As shown in Fig.3, we can see that Tube-P-0-0.5, Tube-P-30-0.5 and Tube-P-180-0.25 are always energetically more favorable than black $\alpha$-P in the temperature range of 0-1700 K. Tube-P-0-0.5 and Tube-P-30-0.5 are always more stable than Tube-P-180-0.25 in the whole temperature range. Interestingly, Tube-P-30-0.5 is energetically more stable than Tube-P-0-0.5 blow 333.29 K and it becomes less stable than Tube-P-0-0.5 when the temperature is higher than 333.29 K.\\
\begin{figure}
\includegraphics[width=3.5in]{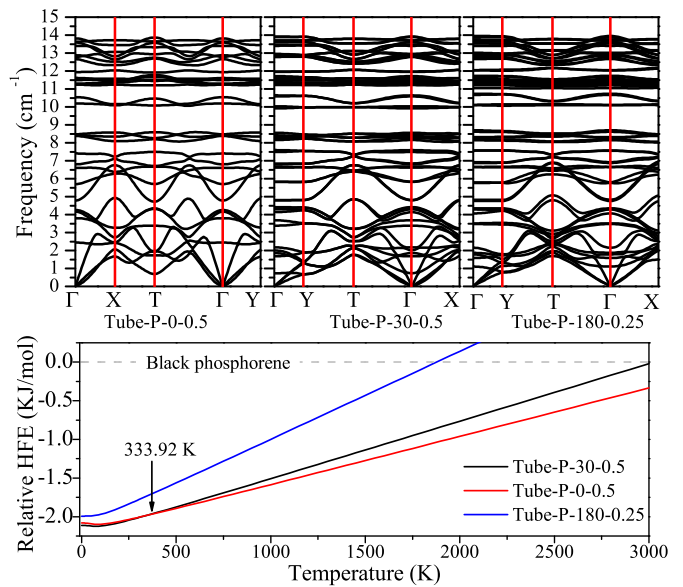}\\
\caption{ The calculated phonon band structures (top) of Tube-P-0-0.5, Tube-P-30-0.5 and Tube-P-180-0.25 as well as their Helmholtz free energies (bottom) relative to that of black $\alpha$-P.}\label{fig3}
\end{figure}
\begin{figure*}
\includegraphics[width=6.0in]{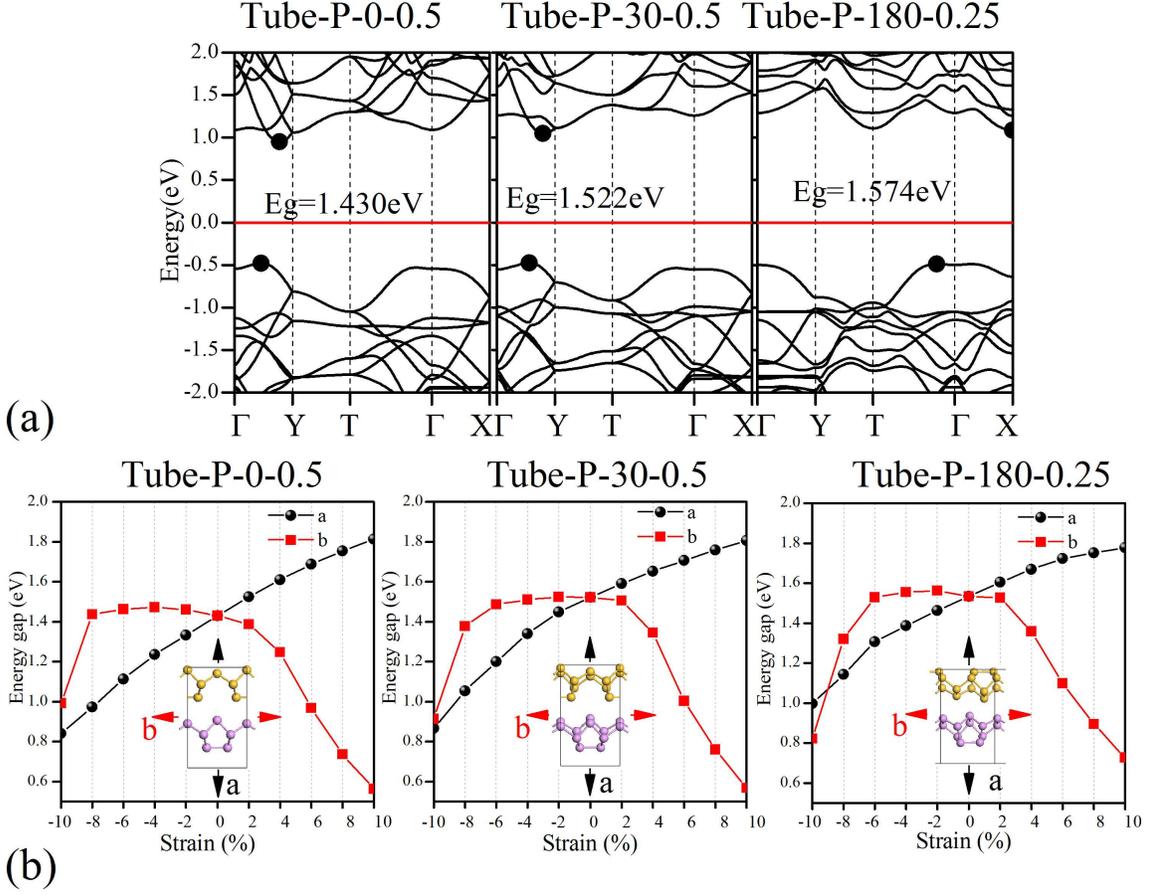}\\
\caption{The calculated electron band structures (a) of Tube-P-0-0.5, Tube-P-30-0.5 and Tube-P-180-0.25, as well as the corresponding modulating effects of stains on their band gaps (b).}\label{fig4}
\end{figure*}
\indent The electronic band structures of allotropes Tube-P-0-0.5, Tube-P-30-0.5 and Tube-P-180-0.25 were investigated by first-principles calculations. As shown in Fig.4, we can see that Tube-P-0-0.5, Tube-P-30-0.5 and Tube-P-180-0.25 are indirect band gap semiconductors with band gaps of 1.430 eV, 1.522 eV and 1.574 eV, respectively. As discussed above, the um-PNT is an indirect band gap semiconductor with a band gap of 2.083 eV, which is larger than those of Tube-P-0-0.5, Tube-P-30-0.5 and Tube-P-180-0.25. That is to say, the inter-tube van der Waals interactions reduce the band gap of the system after assembling. We notice that such a phenomenon is very similar to that of stacking single layers of black $\alpha$-P or blue $\beta$-P into multi-layer \cite{5}. According to such a phenomenon, we can effectively modulate the electronic properties of these new two-dimensional phosphorus allotropes by adjusting the inter-tube distance by strain.\\
\indent As shown in Fig.5 (a), (b) and (c) are the calculated band gaps of Tube-P-0-0.5, Tube-P-30-0.5 and Tube-P-180-0.25 under different strains. We can see that the compressive (tensile) strains cross the tube reduce (increase) the inter-tube distance and correspondingly reduce (increase) the band gap of the systems. Such results are very similar to those in three-dimensional bulk black phosphorus, whose band gap can be effectively modulated by inter-layer distance. The modulating effects of strains along the tube on the band gaps of Tube-P-0-0.5, Tube-P-30-0.5 and Tube-P-180-0.25 are expected similar to that on the single um-PNT where both compressive and tensile strains reduce the band gaps (see in Fig.1 (c)). However, we can see that the results in Fig.5 (a), (b) and (c) are little different to that in Fig.1 (c), especially in the compressive area. For example, in single um-PNT, the band gap always decreases as the compressive strain increases but the band gap of Tube-P-0-0.5 slightly increases at first under small compressive strain and then decreases under larger compressive strain. Such a phenomenon is due to the feature of the inter-tube van der waals interaction in Tube-P-0-0.5, Tube-P-30-0.5 and Tube-P-180-0.25. The compressive strain along the tube first reduce the band gap of the tube itself. Meanwhile it will also increase the inter-tube distance and thus correspondingly increase the band gap of the system. Therefore, when small compressive strain along the tube is applied, the band gap of the system slightly increases.\\
\indent Phosphorene allotropes with lone pairs on the surface like SnS and SnSe \cite{Pi1, Pi2} are expected to be excellent two-dimensional piezoelectric materials \cite{Pi3,Pi4}. However, black $\alpha$-P, blue $\beta$-P and most of all the previously proposed two-dimensional allotropes of phosphorus are centrosymmetric and consequently non-polar, indicating that they are non-piezoelectric. Here, we notice that Tube-P-0-0.5 (Amm2) and Tube-P-30-0.5 (Pmn21) are allowed to be piezoelectric according to their non-centrosymmetric and polar space groups. Their puckered C2v symmetries are very flexible along the direction perpendicular to the tube, which are expected to further enhance the piezoelectricity. Based on the widely used methodology \cite{Pi1, Pi2, Pi5, Pi6} for two-dimensional materials, the piezoelectric coefficients e$_{ijk}^{2D}$ of Tube-P-0-0.5 and Tube-P-30-0.5 are calculated as the third-rank tensors as the relate polarization vector P$_i^{2D}$ to strain $\varepsilon$$_{jk}$ as shown in eq. 1:
\begin{equation}\label{1}
e_{ijk}^{2D}=\frac{\partial P_{i}^{2D}}{\partial \varepsilon_{jk}}
\end{equation}
\indent With mirror symmetry along the tube direction (Y direction), the independent piezoelectric coefficients for Tube-P-0-0.5 and Tube-P-30-0.5 are e$_{111}^{2D}$, e$_{122}^{2D}$ and e$_{212}^{2D}$ = e$_{221}^{2D}$. Indices 1 and 2 correspond to the X and Y directions as indicated in Fig.2 (a). Here, we just consider the responses of polarization P$_1^{2D}$ along X direction to strain $\varepsilon$$_{11}$ and $\varepsilon$$_{22}$ along X and Y directions, respectively. Response of polarization P$_1$$^{2D}$ to share strain $\varepsilon$$_{12}$ is not considered in our calculation. Polarization P$_2^{2D}$ along Y direction is always zero due to the mirror symmetry. Using the Voigt notation \cite{voigt}, we simplify the coefficients as e$_{11}^{2D}$=e$_{111}^{2D}$, e$_{12}^{2D}$=e$_{122}^{2D}$. The piezoelectric coefficients e$_{11}^{2D}$ and e$_{12}^{2D}$ for Tube-P-0-0.5 and Tube-P-30-0.5 are obtained by least-squares fitting of the polarization change per unit area to eq. 2 and eq. 3, respectively.
\begin{equation}\label{2}
P_{1}^{2D}(\varepsilon_{11},\varepsilon_{22}=0)-P_{1}^{2D}(\varepsilon_{11}=0,\varepsilon_{22}=0)=e_{11}^{2D}\varepsilon_{11}
\end{equation}
\begin{equation}\label{3}
P_{2}^{2D}(\varepsilon_{11}=0,\varepsilon_{22})-P_{2}^{2D}(\varepsilon_{11}=0,\varepsilon_{22}=0)=e_{22}^{2D}\varepsilon_{22}
\end{equation}
\begin{figure*}
\includegraphics[width=6.0in]{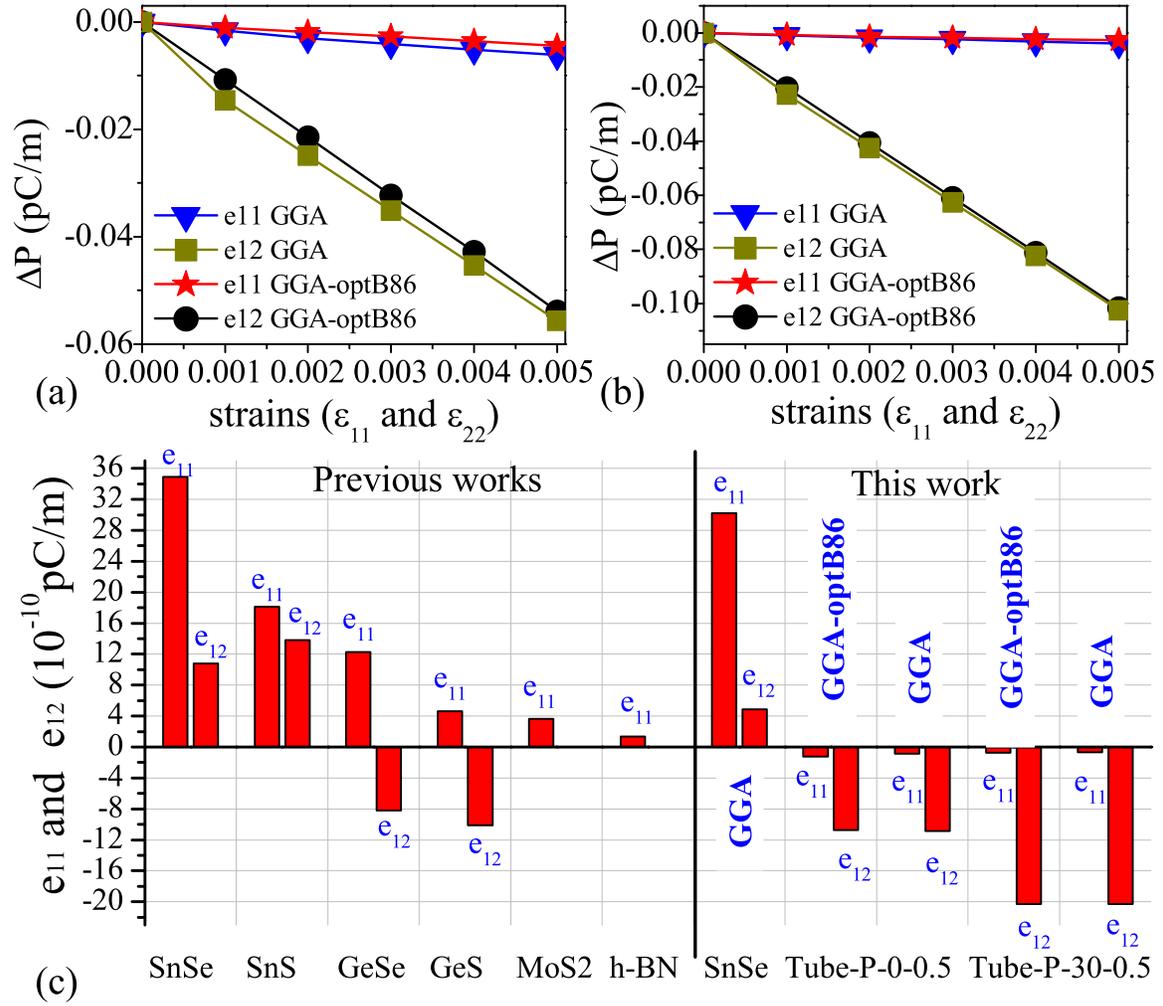}\\
\caption{Polarization changes for  Tube-P-0-0.5 (a) and Tube-P-30-0.5 (b) calculated from GGA and GGA-opt86 methods. Previously calculated piezoelectric coefficients for different two-dimensional materials (left) and the  calculated piezoelectric coefficients for Tube-P-0-0.5 and Tube-P-30-0.5 with/without van der waals function (right) (c).}\label{fig5}
\end{figure*}
\indent In our calculations, we chose the "Relaxed-ion" mode to calculate the change of polarization. Namely, all the atomic positions are fully relaxed under each strain before polarization calculation. We first considered the single layered SnSe as reference system. According to our results, its e$_{11}$ and e$_{22}$ are 30.212 10$^{-10}$C/m and 4.882 10$^{-10}$C/m, respectively. These values are in good agrement with those in previous reports \cite{Pi1, Pi2}. Our calculated results of polarization changes for Tube-P-0-0.5 and Tube-P-30-0.5 are shown in Fig.5 (a) and (b), respectively. After a liner fitting for the polarization change lines, we got the slope for each line, which is considered as the corresponding coefficient. The piezoelectric coefficients of Tube-P-0-0.5 and Tube-P-30-0.5 calculated with/without van der Waals function are shown in Fig.5 (c). Without van der Waals function, the calculated e$_{11}$ and e$_{22}$ for Tube-P-0-0.5 are -0.844 10$^{-10}$C/m and -10.871 10$^{-10}$C/m, respectively. And the calculated e$_{11}$ and e$_{22}$ for Tube-P-30-0.5 are -0.652 10$^{-10}$C/m and -20.311 10$^{-10}$C/m. The consideration of van der Waals function just slightly adjusts these values for Tube-P-0-0.5 and Tube-P-30-0.5. These values are larger than that of MoS$_2$ \cite{Pi3, Pi4, Pi5, Pi6}, indicating that Tube-P-0-0.5 and Tube-P-30-0.5 are good piezoelectric materials for applications in mechanical-electrical energy conversion.\\
\indent In Fig.5 (c), the previously reported piezoelectric coefficients of some well-know 2D materials \cite{Pi1, Pi2, Pi3, Pi4, Pi5, Pi6} are also summarized for comparison. We can see that the piezoelectric coefficients of Tube-P-10-0.5 are comparable to those of the single layered GeSe and GeS and the piezoelectric coefficients of Tube-P-30-0.5 are comparable to those of the single layered SnSe and SnS. These results show that tube-P-0-0.5 and Tube-P-30-0.5 possess remarkable piezoelectricity higher than those of h-BN and MoS$_2$. Especially, Tube-P-0-0.5 and Tube-P-30-0.5 possessing remarkable energetic stability and positive dynamical stability are highly expected to be synthesized in future experiments for piezotronics applications.\\
\indent In summary, based on the previously proposed ultrathin metastable phosphorus nanotube, we predicted a new class of two-dimensional van der Waals crystals for phosphorus through operations of translation and self-rotation. Such a new class of two-dimensional phosphorene allotropes possess remarkable stabilities due to the strong inter-tube van der Waals interactions. They are of about 30-70 meV/atom lower than the single um-PNT depending on the assembling manner. Our calculated results show that most of these two-dimensional van der Waals phosphorene allotropes are energetically more stable than the experimentally viable black $\alpha$-P and blue $\beta$-P. Three of them showing relatively higher probability to be synthesized in future deposition methods were also confirmed to be dynamically stable semiconductors with strain-tunable band gaps. Especially, two of them belonging to polar space groups were predicted possess remarkable piezoelectricity higher than h-BN and MoS$_2$, which may have potential applications for nano-sized sensors, piezotronics, and energy harvesting in portable electronic nano-devices.\\
\indent This work is supported by the National Natural Science Foundation of China (Grant Nos. A040204, 11474244 and 11204261), the National Basic Research Program of China (2015CB921103), the Young Scientists Fund of the National Natural Science Foundation of China (Grant No. 11204260), the Scientific Research Found of HuNan Provincial Education department (No. 14C1095), the Natural Science Foundation of Hunan Province, China (Grant No. 2016JJ3118), and the Program for Changjiang Scholars and Innovative Research Team in University (IRT13093).\\

\end{document}